\documentclass[prb,twocolumn,showpacs]{revtex4}

\usepackage{graphicx}
\usepackage{amssymb}

\newcommand{\la}{\mathord{\langle}}
\newcommand{\ra}{\mathord{\rangle}}

\begin{document}

\title{Scaling behavior of the dipole coupling energy in \\ 
two-dimensional disordered magnetic nanostructures} 

\author{P. J. Jensen} 
\altaffiliation{On leave from: Institut f\"ur 
Theoretische Physik, Freie Universit\"at Berlin, Arnimallee 14, 
D-14195 Berlin, Germany}
\email{jensen@physik.fu-berlin.de}

\author{G. M. Pastor}

\affiliation{
Laboratoire de Physique Quantique, Centre National de la 
Recherche Scientifique, Universit\'e Paul Sabatier, 
F-31062 Toulouse, France } 

{\begin{abstract} 
Numerical calculations of the average dipole-coupling energy
$\overline E_\mathrm{dip}$ in two-dimensional disordered magnetic
nanostructures are performed as function of the particle 
coverage $C$. We observe that $\overline E_\mathrm{dip}$ scales as 
$\overline E_\mathrm{dip}\propto C^{\alpha^*}$ with an unusually small 
exponent $\alpha^*\simeq 0.8$--$1.0$ for coverages $C\lesssim20\%$. 
This behavior is shown to be primarly given by the contributions of
particle pairs at short distances, which is intrinsically related
to the presence of an appreciable degree of disorder. The value of
$\alpha^*$ is found to be sensitive to the magnetic arrangement 
within the nanostructure and to the degree of disorder. For large 
coverages $C\gtrsim20\%$ we obtain 
$\overline E_\mathrm{dip}\propto C^\alpha$
with $\alpha=3/2$, in agreement with the straighforward scaling of the 
dipole coupling as in a periodic particle setup. Taking into account 
the effect of single-particle anisotropies, we show that the scaling 
exponent can be used as a criterion to distinguish between weakly 
interacting ($\alpha^* \simeq 1.0$) and strongly interacting 
($\alpha^* \simeq 0.8$) particle ensembles as function of coverage. 
\end{abstract}


\maketitle 

\section{Introduction}
Nanostructured materials consisting of interacting magnetic 
particles are currently the subject of an intense research activity, 
driven by their fundamental interest and technological 
perspectives. Numerous experimental and theoretical studies have 
been performed for various two-dimensional (2D)\cite{Sun00} and 
three-dimensional (3D)\cite{Luo91,Zha96,Dor99,Lom00} arrangements 
having different degrees of structural and magnetic disorder.
The magnetic behavior of these systems is determined both by 
single-particle properties, like lattice or shape 
anisotropies, and by interparticle interactions, like 
magnetic dipole coupling or indirect exchange. 
The relative importance of these two types of contributions 
can be tuned experimentally, at least to some extent, by varying  
the particle-size distribution, the nanostructure morphology, and 
the average interparticle distance, for example. For the weakly 
interacting regime the single-particle properties dominate, 
and the interactions can be handled as perturbations. However, 
in the most interesting case of strongly interacting systems, where 
the interparticle couplings are stronger than the single-particle 
energies, an explicit treatment of the interactions is crucial. 
Taking into account the complexity of the problem, particularly in the 
presence of disorder, it is a challenge 
to derive simple trends for the dependence 
of the interaction energy on the main nanostructure characteristics.

In this study we focus on the dipole coupling energy $E_\mathrm{dip}$
for magnetic particle ensembles with varying particle coverages $C$ 
and different kinds of nanostructures.\cite{rkky} 
Since the dipole interaction energy between two particles scales 
with the average interparticle distance $R$ as $R^{-3}$, and since 
$C$ is propotional to $R^{-D}$, $D$ being the 
spatial dimension, one could in principle expect that the average 
dipole energy $\overline E_\mathrm{dip}(C)$ should be proportional to 
$C^\alpha$, with $\alpha=3/D$. However, several previous studies 
indicate a different scaling behavior for disordered dipole-coupled 
systems.\cite{Zha96,Dor99,Lom00} For example, a weaker decrease of 
$\overline E_\mathrm{dip}$ characterized by an \textit{effective exponent} 
$\alpha^*=0.8\pm0.1$ has been found experimentally upon thinning of 
ferrofluids ($D=3$).\cite{Zha96} In addition, the strength of the dipole 
coupling seems to be underestimated as compared to measurements, when 
one calculates its effect on the energy barrier for single-particle 
magnetic reversals.\cite{Dor99} Moreover, numerical results for 
$\overline E_\mathrm{dip}$ of a ferrofluid monolayer ($D=2$) with 
different particle coverages reveal significant deviations from the 
$\alpha=3/2$ scaling.\cite{Lom00} These experimental and theoretical 
findings suggest that the effective dipole energy in disordered
systems decreases more slowly with decreasing coverage than originally 
expected. 

The purpose of this study is to determine the role of global 
nanostructure characteristics, such as coverage and degree of disorder, 
on the average dipole-coupling energy of 2D particle ensembles. In 
particular we show that the scaling exponent $\alpha$, 
$E_\mathrm{dip}\propto C^\alpha$, allows to discern whether $E_\mathrm{dip}$ 
is dominated by the average interparticle distance $R$, as in a 
periodic particle setup, or by short-distance contributions. Moreover, 
taking into account single-particle anisotropies we demonstrate that 
the value of $\alpha$ can be used as a criterion to distinguish
between weakly and strongly interacting disordered particle ensembles. 

\section{Model Simulations}
The 2D nanostructure is modeled using a square unit cell containing 
non-overlapping, sphere-shaped magnetic particles, and periodic
boundary conditions. We consider disordered particle ensembles with a
fully random distribution within the unit cell, as well as disturbed 
square arrays where the particles are scattered around periodic
lattice sites by applying a Gaussian distribution $P(\mathbf{r})$ with 
standard deviation $\sigma_R$. For such planar particle ensembles the 
overall coverage $C$ is related to the average interparticle distance 
$R>2\,r_0$ by $C=\pi\,(r_0/R)^2$. A uniform-particle size distribution 
with $N\simeq24000$ atoms per particle is assumed, which corresponds
to a particle radius $r_0\simeq16\,a_0$, where $a_0$ is the 
nearest-neighbor interatomic distance. 
Due to the strong interatomic exchange interaction and for particle 
diameters smaller than typical magnetic domain wall widths, each 
particle can be regarded as a single-domain Stoner-Wohlfarth 
magnet\cite{StW48} carrying a giant spin $M\simeq N\,\mu_{at}$, 
where $\mu_{at}$ is the magnetic moment per atom. 
For simplicity, the particle magnetizations $\mathbf{M}_i$ are 
restricted to be within the plane of the nanostructure. The magnetic 
configuration is thus characterized by the set of in-plane angles 
$\{\phi_i\}$, with $i=1,\ldots,N$. For each given particle distribution 
and magnetic configuration we determine the magnetic energy 
\begin{eqnarray} 
E &=& E_\mathrm{anis} + E_\mathrm{dip} = -K_2\sum_i 
(\mathbf{M}_i\,\mathbf{n}_i)^2 \nonumber \\  
&&\hspace*{-1cm} + \frac{1}{2}\;\sum_{i,j \atop i\ne j} 
\left[\mathbf{M}_i\;\mathbf{M}_j\;r_{ij}^{-3}-
3\big(\mathbf{r}_{ij}\,\mathbf{M}_i\big)
\big(\mathbf{r}_{ij}\,\mathbf{M}_j\big)\;r_{ij}^{-5}\right]  
\;. \label{e3} \end{eqnarray} 
The first term represents the single-particle anisotropy energy of 
particle $i$, resulting from spin-orbit interactions and the shape 
anisotropy. For simplicity, a uniaxial anisotropy is assumed with 
strength $K_2$ per atom and easy axis directed along a given unit 
vector $\mathbf{n}_i$. The second term of Eq.(\ref{e3}) is the 
dipole-dipole interaction between magnetic moments $\mathbf{M}_i$, 
where $r_{ij}=|\mathbf{r}_{ij}|=|\mathbf{r}_i-\mathbf{r}_j|$ is 
the distance between the centers of particles $i$ and $j$.  
The infinite range of the dipole interaction is taken into account 
by applying an Ewald-type summation over all periodically arranged 
unit cells of the extended planar system.\cite{Jen97} 
In addition, we also include the leading correction resulting 
from the dipole-quadrupole interaction, which takes 
into account the effect of finite particle sizes.\cite{JeP02} 

In order to determine the average magnetic energy 
$\overline E=\overline E_\mathrm{anis}+\overline E_\mathrm{dip}$ 
we start from an arbitrary initial guess of the angles $\phi_i$ of 
each particle, and relax the magnetic configuration to a nearby 
local minimum of the total energy, using a conjugated gradient 
method. The resulting relaxed states depend in general on the initial
configurations and show strong magnetic noncollinearities. The large 
diversity of 
metastable states which is characteristic of disordered particle 
arrangements\cite{Luo91,Dor99,JeP02} is taken into account by 
averaging $E$ over many different random initial guesses 
(typically $\sim100$) for the same spatial arrangement of the 
particles. Notice that this procedure is not intended to
search preferently for the global energy minimum. 
For comparison, $\overline E$ is also calculated for the same particle 
setup but assuming a parallel alignment of all particle magnetizations 
(i.e., a ferromagnetic state with $\phi_i=\phi_\mathrm{fm}$)  
and averaging over the common magnetization angle $\phi_\mathrm{fm}$. 
Note that for a disordered particle ensemble such a ferromagnetic
state does not usually correspond to a local energy minimum. In
addition, a second average over the spatial configuration of the
nanostructure is necessary in order to minimize spurious 
effects of the finite size of the unit cell. For this purpose, 
an appropriate number (typically $\sim200$) of different realizations 
of the unit cell is considered, using the same global variables which 
characterize the particle ensemble. In the following calculations 
we use parameters appropriate for Fe, i.e., $\mu_{at}=2.2\;\mu_B$ and 
$a_0=2.5$~\AA, yielding the dipole energy unit 
$E_\mathrm{dip}^0=\mu_{at}^2/a_0^3=0.19$~K. Concerning the strength $K_2$ 
of the uniaxial anisotropy we consider values appropriate for films 
and particles of 3d transition metals ($K_2=0.1-1$~K/atom).\cite{BlH94} 
\begin{figure}[t] 
\includegraphics[width=8cm]{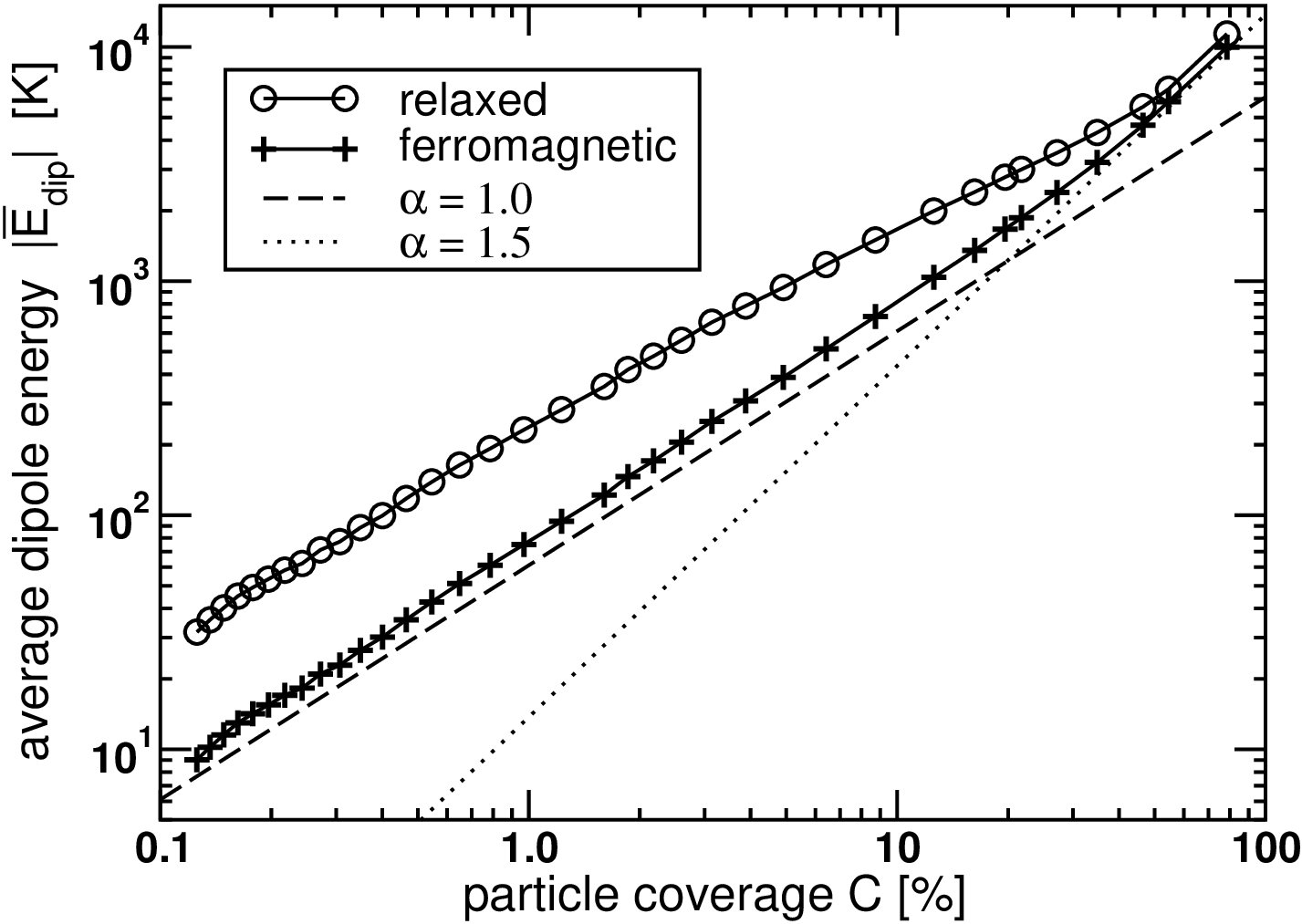} 
\caption{Average dipole energy $|\overline E_\mathrm{dip}(C)|$ per 
particle of random 2D ensembles of magnetic particles as function of
the particle coverage $C$, neglecting single-particle anisotropies 
($K_2=0$). Results are given for relaxed magnetic states (circles),
and for ferromagnetic ensemble magnetizations (crosses). The straight 
lines correspond to $\overline E_\mathrm{dip}(C)\propto C^\alpha$,
with $\alpha=1$ (dashed) and $\alpha=3/2$ (dotted). }\end{figure} 

\section{Results and Discussion}
First we focus on strongly interacting 2D systems of magnetic 
particles, neglecting the single-particle contribution ($K_2=0$). 
In Fig.1 results are given for the absolute value of the 
average dipole energy per particle $|\overline E_\mathrm{dip}(C)|$
as function of particle coverage $C$ for a fully random particle 
ensemble. $|\overline E_\mathrm{dip}(C)|$ is shown for the relaxed 
magnetic arrangements as well as for the ferromagnetic ones. 
From the log-log-plot of $|\overline E_\mathrm{dip}(C)|$ 
one observes that the exponent $\alpha=3/2$, corresponding to the 
straightforward scaling of the dipole interaction for $D=2$, is 
approximately valid only for large coverages $C\gtrsim 20\,\%$. For 
smaller $C$, in the investigated range 
$0.1\%\le C\lesssim 20\,\%$, one finds a different dependence with 
clearly smaller effective exponents. For the ferromagnetic arrangement 
we obtain $\alpha_\mathrm{fm}^*\simeq1.0$ and, quite interestingly, 
for the relaxed solution an even smaller value 
$\alpha_\mathrm{r}^*\simeq0.8-0.9$. The numerical accuracy of the 
calculated exponents $\alpha_\mathrm{fm}^*$ and $\alpha_\mathrm{r}^*$ 
is estimated to be about $\pm0.02$. 

These remarkable results can be qualitatively explained by the 
following simple analysis. Consider two non-overlapping particles with 
radius $r_0$ in a $D$-dimensional space coupled by an interaction 
$E_n(r)\propto r^{-n}$. Let the first particle be located 
at $r=0$, and let $P(r)$ be the probability to find the second one 
between $2\,r_0$ and a maximal distance $R_\mathrm{max}$. 
The upper bound $R_\mathrm{max}$ is related to the particle 
concentration by $C\propto(r_0/R_\mathrm{max})^D$, whereas the 
lower bound $2\,r_0$ results from the hard core (excluded volume) 
of the non-overlapping particles. 
The average interaction energy $\overline E_n$ 
of such a particle pair is proportional to the average 
\begin{equation} 
\left\langle r^{-n} \right\rangle 
= \int_{2r_0}^{R_\mathrm{max}} dr\,r^{D-1} \, P(r) \, r^{-n} 
\bigg/ \int_{2r_0}^{R_\mathrm{max}} dr\,r^{D-1} \, P(r) \,.
\label{e1} \end{equation} 
For a random distribution corresponding to a constant $P(r)$ one obtains 
\begin{equation}
\la r^{-n}\ra=
\frac{D}{n-D}\;\frac{R_\mathrm{max}^{n-D}-(2r_0)^{n-D}}
{(2r_0R_\mathrm{max})^{n-D}\,[R_\mathrm{max}^D-(2r_0)^D]}  \label{e2a}
\end{equation} 
for $n>D$, and 
\begin{equation}
\la r^{-n}\ra=
D\;\frac{\ln(R_\mathrm{max}/2r_0)}{R_\mathrm{max}^D-(2r_0)^D} 
\label{e2b}  \end{equation} 
for $n=D$. In the limit of low coverages, 
$2\,r_0\ll R_\mathrm{max}\;\to\;\infty$, this yields 
\begin{equation}  \la r^{-n}\ra
\; \simeq \; \frac{D}{n-D}\;\frac{2^D}{(2r_0)^n}
\bigg(\frac{r_0}{R_\mathrm{max}}\bigg)^D 
\; \propto \; (2r_0)^{-n}\; C \;, \label{e2c} 
\end{equation} 
for $n>D$, and a similar limit for $n=D$. For $n<D$ one finds 
$\la r^{-n}\ra\propto C^{n/D}$, which coincides with 
straightforward scaling. Notice that for $n\ge D$ the average 
interaction energy $\overline E_n(R_\mathrm{max})$ for 
$R_\mathrm{max}\to\infty$ is no longer proportional to 
$(R_\mathrm{max})^{-n}$. Instead, an effective exponent $n^*=D$ 
is obtained. Consequently, as function of concentration one obtains
$\overline E_n(C)\propto C^{\alpha^*}$ with $\alpha^*=1$. 
The physical reason behind this behavior is that only the upper bound 
$R_\mathrm{max}$ scales with $C$, since it is proportional to the 
average interparticle distance, but not the lower bound $2\,r_0$. 
Hence, the \textit{average} energy is dominated by 
the interactions between particle pairs at distances of the order 
of the particle diameter, which is proportional to $(2\,r_0)^{-n}$,  
multiplied by the probability to find such a particle pair, which is 
proportional to $C$. In other words, as a consequence of the 
relatively strong distance dependence of the interaction for 
$n\ge D$, in a diluted random system the average energy 
$\overline E_n(C)$ is dominated by the short-distance contributions, 
and not by the energy corresponding to the average interparticle 
distance. Notice that if $n\ge D$, 
$\overline E_n\propto C$ is independent of the particular kind of 
interaction characterized by the number $n$. Moreover, this behavior 
of a random particle ensemble is also obtained for an exponentially 
decreasing interaction $E\propto\exp(-\lambda\,r)$ for all $D$. 

In the context of the present study corresponding to $D=2$ and $n=3$, 
one concludes that a 2D \textit{random} ensemble of magnetic 
nanoparticles exhibits a stronger than expected average dipole 
interaction, which scales with the particle coverage as 
$\overline E_\mathrm{dip}(C)\propto C^{\alpha^*}$ with 
$\alpha^*\simeq1.0$. In contrast, if one varies the interparticle 
distance in a \textit{periodic} array the smallest distance between 
particle pairs also scales with coverage, yielding therefore the 
straightforward scaling behavior with $\alpha=3/D$. 
Thus, by analyzing the exponent $\alpha$ one can distinguish 
whether the average dipole energy is dominated by short-distance 
contributions or by the energy corresponding to the average 
interparticle distance. The difference between $\alpha$ and 
$\alpha^*$ is significant for 1D and 2D systems, whereas for 3D 
systems one expects $\alpha\simeq\alpha^*\simeq1$. 

Another interesting effect is the difference in the exponents 
$\alpha_\mathrm{r}^*$ and $\alpha_\mathrm{fm}^*$ obtained for the 
relaxed and ferromagnetic arrangements. This seems to be an indication 
of the formation of particle chains with correlated 'head-to-tail' 
alignments of the magnetizations of closeby particles. Such magnetic 
rearrangements are possible only by relaxing the angles $\phi_i$. In 
fact, recent calculations have shown that $|\overline E_\mathrm{dip}|$ 
increases with increasing positional disorder as a result of magnetic 
relaxations.\cite{JeP02} Moreover, reduced values of
$\alpha^*=0.8\pm0.1$ have also been observed in experiments 
on frozen high-density ferrofluids.\cite{Zha96,Cha96} 

By closer inspection of Fig.1 one observes that the exponent 
$\alpha_\mathrm{r}^*$ is not exactly the same for all coverages, 
since a slight change of slope appears in the curve of the relaxed
solution. Moreover, differences between 
$\alpha_\mathrm{r}^*$ and $\alpha_\mathrm{fm}^*$ are found in 
the coverage range where magnetic relaxation driven by the dipole 
coupling is significant, i.e., in the strongly interacting regime. 
This can be demonstrated by considering in 
addition to the dipole coupling a single-particle uniaxial anisotropy 
$K_2$ with randomly distributed directions $\mathbf{n}_i$ of the easy 
axes. Unless the average anisotropy energy $|\overline E_\mathrm{anis}|$ 
is so strong that it dominates over 
$|\overline E_\mathrm{dip}|$ even for large coverages, 
a crossover from the strongly interacting regime 
($|\overline E_\mathrm{anis}|<|\overline E_\mathrm{dip}|$) 
to the weakly interacting regime 
($|\overline E_\mathrm{anis}|>|\overline E_\mathrm{dip}|$) takes 
place with decreasing $C$ at the \textit{crossover coverage} 
$\tilde C(K_2)$. This is shown in Fig.2 where, 
as in Fig.1, $|\overline E_\mathrm{dip}|$ and 
$|\overline E_\mathrm{anis}|$ are given as function of $C$ for
different values of $K_2$. We have determined the relaxed magnetic 
arrangements in presence of local anisotropies, and have derived the 
scaling exponents $\alpha^*$ corresponding to the dipole contribution 
of the magnetic energy. First of all, one observes that the scaling 
behavior, as estimated by Eq.(\ref{e2a}) and depicted in Fig.1, 
is also valid in presence of single-particle anisotropies. In the 
coverage range where $|\overline E_\mathrm{dip}|$ is dominated by 
short-distance contributions, i.e., for $C\lesssim20\%$, 
and in the strongly interacting regime $C>\tilde C$ where 
$|\overline E_\mathrm{dip}|>|\overline E_\mathrm{anis}|$, a reduced 
value of $\alpha_\mathrm{r}^*<1$ is clearly observed. As $C$ is 
further decreased, $\alpha_\mathrm{r}^*$ changes gradually to 
$\alpha_\mathrm{r}^*\simeq1.0$ below the crossover coverage 
$\tilde C$. This can be understood by recalling that magnetic 
relaxation driven by dipole coupling is significant only in the 
strongly interacting regime. In the opposite case of weak interactions 
($C<\tilde C$) relaxation due to dipole coupling is negligible, thus 
resulting in $\alpha_\mathrm{r}^*\simeq\alpha_\mathrm{fm}^*\simeq1.0$. 
Note that for the largest considered value of $K_2$ the crossover 
happens directly from $\alpha=3/2$ to $\alpha_\mathrm{r}^*\simeq1.0$, 
since here $\tilde C(K_2)\simeq20\%$. Consequently, the determination 
of the scaling exponent $\alpha_\mathrm{r}^*$ allows one to discern 
between strongly and weakly interacting ensembles. 
\begin{figure} 
\includegraphics[width=8cm]{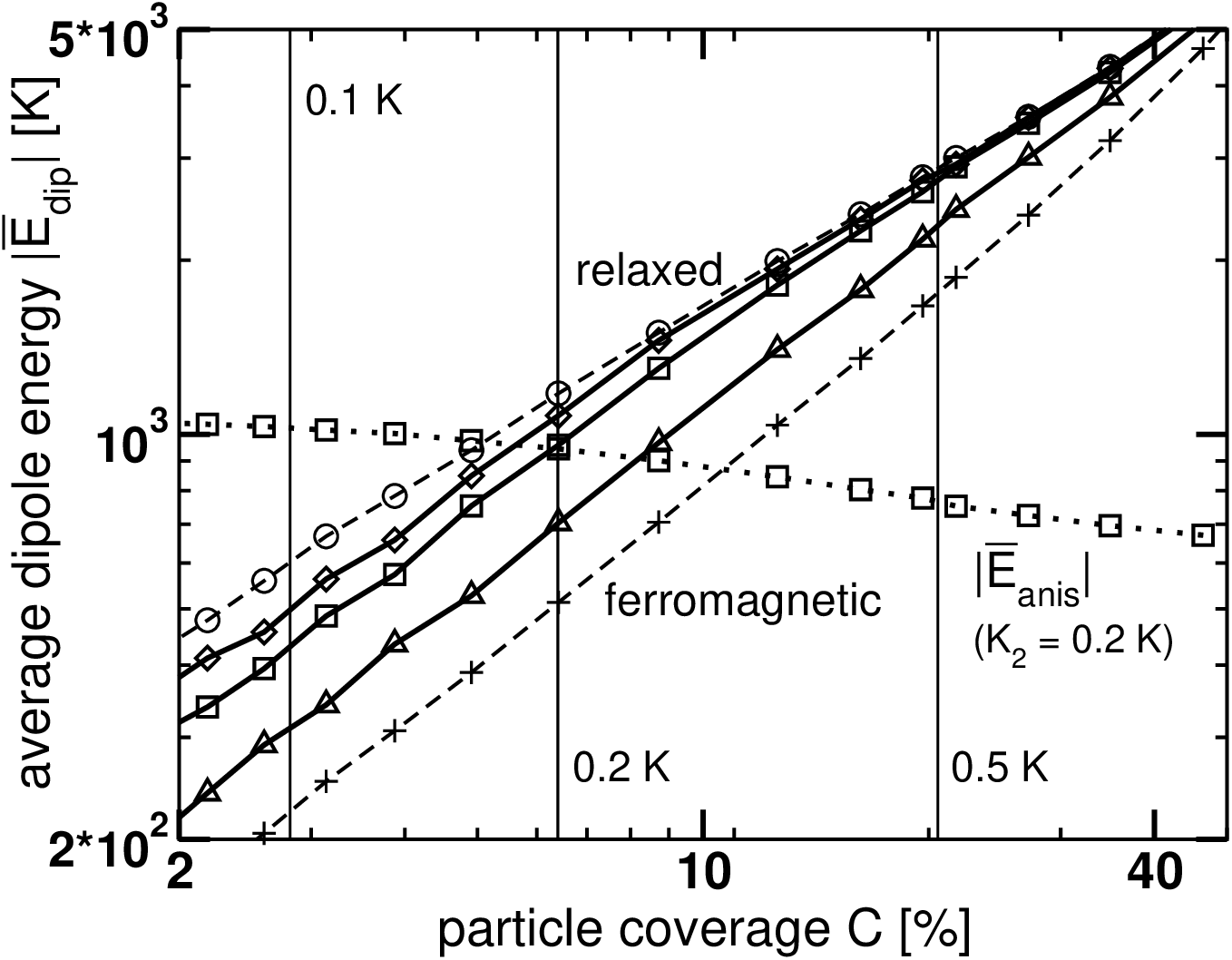} 
\caption{Average dipole energy $|\overline E_\mathrm{dip}(C)|$ 
for the relaxed magnetic states as function of the particle 
coverage $C$ for several values of the single-particle 
uniaxial anisotropy $K_2=0.1$~K/atom (diamonds), $K_2=0.2$~K/atom 
(squares), and $K_2=0.5$~K/atom (triangles). For $K_2=0.2$~K/atom, 
the average anisotropy energy $|\overline E_\mathrm{anis}(C)|$ is also
shown (dotted line, squares). The vertical lines indicate the 
anisotropy-dependent crossover coverages $\tilde C(K_2)$, 
for which $\overline E_\mathrm{dip}=\overline E_\mathrm{anis}$. 
For the sake of comparison the dashed curves show 
$|\overline E_\mathrm{dip}(C)|$ for $K_2=0$ for both the relaxed 
states (circles) and ferromagnetic states (crosses) as in Fig.1. 
}\end{figure} 

The average anisotropy energy $|\overline E_\mathrm{anis}|$ is of
course independent of coverage $C$ in the absence of dipole interactions 
and other magnetic couplings. A dependence on $C$ appears only if the 
interactions force the particle magnetizations $\mathbf{M}_i$ to
deviate from the local easy axes $\mathbf{n}_i$, cf.\ Eq.(\ref{e3}).  
Moreover, the exponent $\alpha_\mathrm{fm}^*$ for 
a ferromagnetic state is not affected at all by single-particle 
anisotropies. Since for such an arrangement a magnetic relaxation is not 
effective, the single-particle energy yields a contribution to the total 
magnetic energy which is independent of the coverage. For anisotropies 
with aligned easy-axes directions we obtain the same behavior of 
$\alpha_\mathrm{r}^*$ as for randomly distributed ones, since the
dipole-coupling-induced relaxation of the particle magnetizations is
not much affected by the easy axes distribution. 

In order to study the differences between $\alpha_\mathrm{r}^*$ and 
$\alpha_\mathrm{fm}^*$ in further detail we have also considered 1D 
systems of dipole-coupled particles. In this case the particle 
magnetizations are aligned parallel to the chain axes in the most 
stable state. A numerical investigation of such 1D systems with
randomly assembled particles yields effective exponents 
$\alpha_\mathrm{fm}^*\simeq\alpha_\mathrm{r}^*\simeq1.0$. This is
quite reasonable since in this case the relaxed and ferromagnetic 
solutions coincide. In the case of 1D stripes with finite widths, we  
observe a smooth crossover from $\alpha_\mathrm{r}^*\simeq1$ to 
$\alpha_\mathrm{r}^*\simeq0.8$ with increasing stripe width
in the investigated coverage range $0.1\%\le C\lesssim 20\,\%$. These 
findings confirm the idea that the reduction of $\alpha_\mathrm{r}^*$ 
with respect to $\alpha_\mathrm{fm}^*$ in 2D and 3D nanostructures is 
the result of magnetic relaxations. 
\begin{figure} 
\includegraphics[width=8cm]{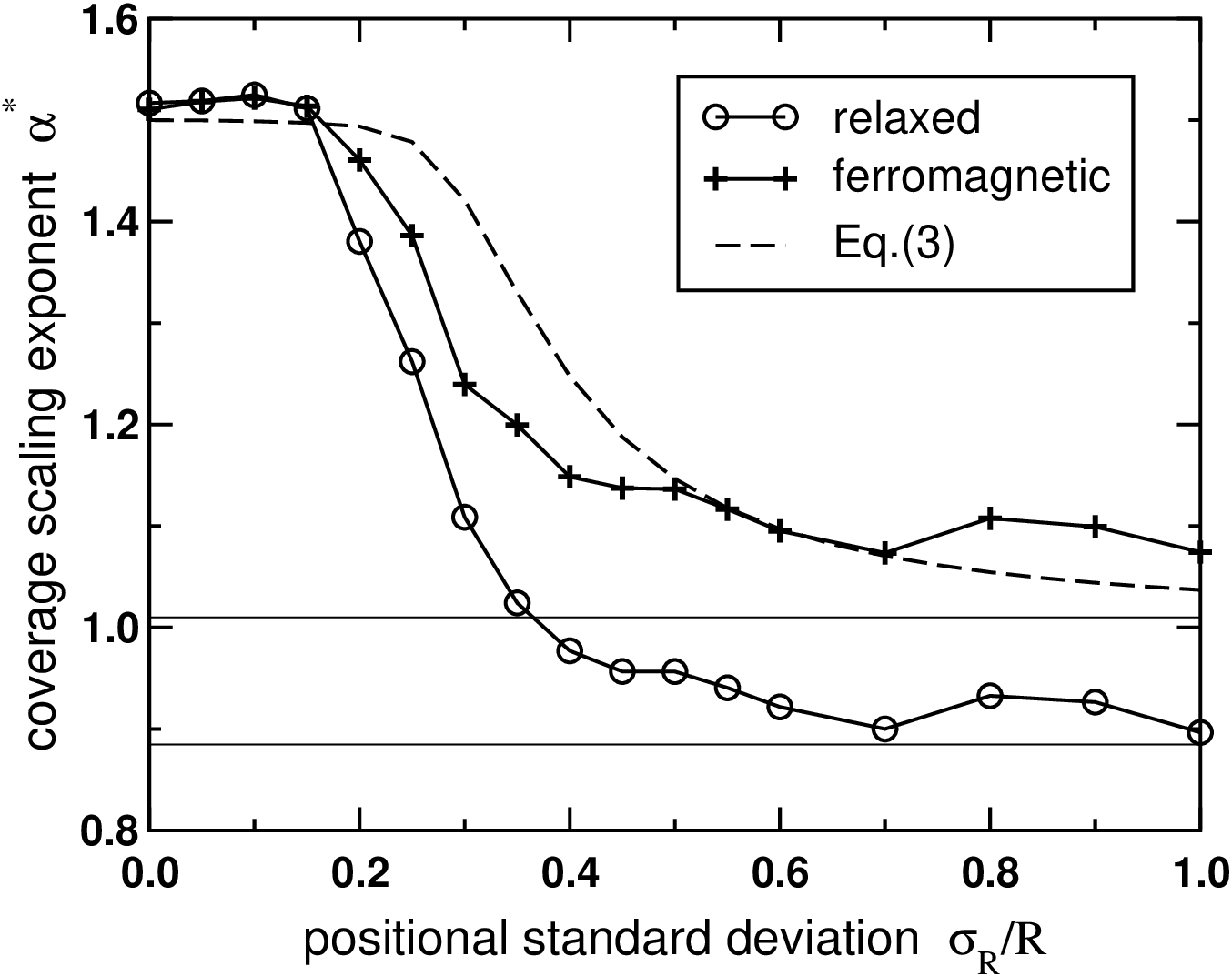} 
\caption{Scaling exponents $\alpha_\mathrm{r}^*$ and  
$\alpha_\mathrm{fm}^*$ of the average dipole energy 
$|\overline E_\mathrm{dip}(C)|\propto C^\alpha$ corresponding 
to the relaxed and ferromagnetic arrangements as function of the 
positional standard deviation $\sigma_R$ in units of the average 
interparticle distance $R$. A disturbed square particle array is 
assumed, neglecting single-particle anisotropies ($K_2=0$). The 
exponents correspond to the coverage range $0.1\%<C<20\%$. The dashed 
curve refers to $\alpha^*$ as derived from Eq.(\ref{e2a}). The 
horizontal lines indicate $\alpha_\mathrm{r}^*$ and 
$\alpha_\mathrm{fm}^*$ calculated for a fully random particle
ensemble. }\end{figure} 

To quantify the role of disorder on the coverage dependence 
of $\overline E_\mathrm{dip}$ we have studied the crossover 
between periodic and fully random particle setups by 
considering systems with intermediate degrees of positional 
disorder. Here the single-particle anisotropy is neglected for 
simplicity ($K_2=0$). Two-dimensional particle ensembles have been 
constructed with a quasi-periodic arrangement in which 
the particles are scattered around the sites of a square lattice 
following a Gaussian-like distribution with standard deviation 
$\sigma_R$. Thus, a  
periodic square lattice corresponds to $\sigma_R=0$, and the  
particle arrangement becomes increasingly disordered 
with increasing $\sigma_R$. For each  $\sigma_R$ 
the average dipole energy $\overline E_\mathrm{dip}(C)$ 
has been determined as function of particle coverage $C$.  
The corresponding effective exponents $\alpha_\mathrm{r}^*$ and 
$\alpha_\mathrm{fm}^*$ for the relaxed and ferromagnetic 
states are presented in Fig.3 as function of $\sigma_R/R$, 
where $R$ is the average interparticle distance. For small 
$\sigma_R$, i.e., $0<\sigma_R/R\lesssim0.2$, one observes that 
$\alpha_\mathrm{r}^*\simeq\alpha_\mathrm{fm}^*\simeq3/2$, in 
agreement with the straightforward scaling of the dipole 
coupling. As $\sigma_R$ increases beyond $\sigma_R/R=0.2$, 
the exponents decrease rapidly, approaching the limit 
of a fully random particle ensemble. In fact, previous 
calculations have shown that $\overline E_\mathrm{dip}(\sigma_R)$ 
is close to $\overline E_\mathrm{dip}$ of random particle 
arrangements already for $\sigma_R/R=0.4\,-\,0.6$.\cite{JeP02} 
These numerical results are corroborated by the simple analytical 
estimate given by Eq.(\ref{e2a}), using for $P(\mathbf{r})$ a Gaussian 
distribution with mean $R_\mathrm{max}/2$ and standard deviation 
$\sigma_R$. The different scaling properties of ordered and disordered 
nanostructures confirm the importance of the sample 
characteristics on the effective scaling exponents. 

Finally, several other effects have been investigated which could in 
principle modify the values of the exponents $\alpha^*$. We have 
considered (i) unit cells with different numbers of particles, (ii) 
particle-size dispersions with Gaussian or log-normal distributions,   
(iii) disk-shaped particles with the same particle radius $r_0$ as 
the previously considered sphere-shaped particles, and (iv) the role 
of dipole-quadrupole corrections to the point-dipole sums. 
While the magnitude of $\overline E_\mathrm{dip}$ is obviously 
affected by some of these contributions, 
in all cases the effective exponents $\alpha_\mathrm{r}^*$ and 
$\alpha_\mathrm{fm}^*$ remain unchanged within numerical accuracy. 
These results confirm the validity of the previously discussed trends
and reinforce the idea that $\alpha_\mathrm{r}^*$ can be used as a
parameter to characterize some aspects of the magnetic behavior of
interacting magnetic nanostructures with varying degrees of disorder. 

\section{Conclusion}
The average dipole energy $E_\mathrm{dip}$ of two-dimensional 
disordered magnetic nanostructures has been determined by using
numerical simulations and simple analytical estimates. 
A coverage dependence of the form  
$|\overline E_\mathrm{dip}(C)|\propto C^{\alpha^*}$ is obtained with
nearly constant $\alpha^*$ over a wide coverage range. For 
$0.1\%<C<20\%$ the exponent $\alpha^*\simeq0.8$--$1.0$ is
significantly smaller than the value $\alpha=n/D=3/2$ expected from 
a straightforward scaling of the dipole interaction. This scaling 
results from dominating short-distance contributions, and is 
intrinsically related to the presence of significant disorder. 
In contrast, for $C\gtrsim20\%$ the behavior of 
$\overline E_\mathrm{dip}(C)$ is determined mainly by the average 
interparticle distance, which depends on coverage as 
$R\propto C^{-1/D}$, and therefore yields $\alpha=3/2$, as for a 
periodic particle setup. A smooth transition between these two scaling 
regimes is observed with increasing degree of disorder. 

Taking into account the effect of single-particle anisotropies, 
we have shown that the value of the scaling exponent $\alpha^*$ 
allows to distinguish between strongly and weakly interacting particle 
ensembles, since $\alpha^*$ is sensitive to the magnetic 
arrangement within the nanostructure. Indeed, one observes 
a crossover from $\alpha^*\simeq0.8$, when magnetic relaxations 
driven by dipole couplings dominate, to $\alpha^*\simeq1.0$, 
when the directions of the particle moments are determined mainly by 
local anisotropies. It would be very interesting to extend 
these investigations to 3D nanostructures in view of the numerous 
experimental results that are available for such systems. In addition, 
the scaling behavior should be analyzed as function of temperature, 
since it is likely to be affected by thermally induced crossings of 
energy barriers between different magnetic arrangements. 

We would like to thank Professor Dr.\ K.H.\ Bennemann for useful 
discussions. Support from the EU GROWTH project AMMARE (contract 
number G5RD-CT-2001-00478), and from the Deutsche 
Forschungsgemeinschaft, SFB 450, TP C4, is gratefully acknowledged. 
Computer resources were provided by IDRIS (CNRS, France).


\begin{thebibliography}{99}
\bibitem{rkky} In addition a strong interaction in ensembles of 
magnetic particles might be caused also by the long-range indirect 
exchange. For example, the conduction electrons of a metallic support 
yield the oscillating Ruderman-Kittel-Kasuya-Yoshida (RKKY-) coupling.
\end{thebibliography}
\end{document}